\documentclass[english,prd,twocolumn,superscriptaddress,nofootinbib,preprintnumbers]{revtex4-1}
\usepackage[latin1]{inputenc}
\usepackage{graphicx}
\usepackage{amssymb}
\usepackage{color}
\usepackage{float}
\usepackage{amsmath}
\usepackage{amsfonts}
\usepackage{dcolumn}
\usepackage{hyperref}
\usepackage{subfigure}
\usepackage{mathtools}
\usepackage{soul,xcolor}

\newcommand{\be}{\begin{equation}}
\newcommand{\ee}{\end{equation}}
\newcommand{\bea}{\begin{eqnarray}}
\newcommand{\eea}{\end{eqnarray}}

\begin{document}

\title{Pure Lovelock black hole in the dimension, $d=3N+1$, is stable}

\author{Radouane Gannouji}
\affiliation{Instituto de F\'{\i}sica, Pontificia Universidad  Cat\'olica de Valpara\'{\i}so, Casilla 4950, Valpara\'{\i}so, Chile}
\author{Yolbeiker Rodr\'{\i}guez Baez}
\affiliation{Universidad T\'{e}cnica Federico Santa Mar\'{\i}a, Casilla 110-V, Valpara\'{\i}so, Chile}
\affiliation{Instituto de F\'{\i}sica, Pontificia Universidad  Cat\'olica de Valpara\'{\i}so, Casilla 4950, Valpara\'{\i}so, Chile}
\author{Naresh Dadhich}
\affiliation{IUCAA, Post Bag 4, Ganeshkhind, Pune 411 007, India}

\begin{abstract}
In this paper we show that pure Lovelock static Schwarzschild's analogue black hole in dimensions $d>3N+1$, where $N$ is the degree of Lovelock polynomial action, is stable even though pure Gauss-Bonnet $N=2$ black hole is unstable in dimension $d<7$. We also discuss and compare quasinormal modes for pure Lovelock and the corresponding Einstein black hole in the same dimension. We find that perturbations decay with characteristic time which is weakly dimensional dependent as it depends only on the gravitational potential of the background solution, while frequency of oscillations however depend on the dimension. Also we show that spectrum of perturbations is not isospectral except in $d=4$.
\end{abstract}

\maketitle

\section{Introduction}

It is well known static black hole described by Schwarzschild's vacuum solution of the Einstein
equation is stable relative to metric perturbations in all dimensions $\geq4$ \cite{Ishibashi:2003ap}. In this
case gravitational potential goes as $1/r^{d-3}$ with $d-3\geq1$ always. In other words, so long as
potential falls sharper than $1/r$, it would be stable. The natural question that
arises is, would it also be so for static black hole in other generalizations of Einstein's gravity?
The most natural and interesting generalization of Einstein's gravity -- general relativity (GR) is
the Lovelock theory involving higher order derivatives as its action is homogeneous polynomial in
Riemann curvature of arbitrary degree $N$. Despite this, it is remarkable that the equation that
follows from variation of the action always yields second order equation of motion.

Einstein gravity is included in this generalized theory for $N=1$ linear order in Riemann while
$N=2$ is the quadratic Gauss-Bonnet, and so on. Each order comes with a dimensionful coupling constant,
and there is sum over $N = 0,1, 2, ...$ where $N=0$ corresponds to cosmological constant, $\Lambda$.
The general feature of the Lovelock theory is that for $d=2N$ equation is vacuous meaning corresponding
Einstein tensor $G_{ab}$ vanishes identically, it is kinematic for $d=2N+1$ implying corresponding
Riemann is entirely given in terms of Ricci and hence there can exist no non-trivial vacuum solution
\cite{dadhich:2012}. Gravity could have dynamics admitting non-trivial vacuum solution only in dimensions
$d\geq2N+2$. This means, this is a quintessentially higher dimensional generalization; i.e. the
term corresponding to $N>1$ will make non-zero contribution to equation of motion only in dimension
greater than four.

By defining Lovelock analogue of Riemann curvature \cite{dadhich:2008,kastor:2012,dadhich:2012}, it has been shown that
gravity is kinematic -- corresponding Riemann given in terms of Ricci -- in all critical odd dimensions,
$d=2N+1$ \cite{camanho:2015} for pure Lovelock Lagrangian that includes only one $N$th order term. By pure Lovelock we mean the Lagrangian and consequently equation of motion has only one $N$th
order term, there is no sum over lower orders. In particular pure Gauss-Bonnet has only the Gauss-Bonnet
term without Einstein-Hilbert term. Apart from kinematicity property being universal for all critical
odd $d=2N+1$ dimensions, pure Lovelock is also singled out by another property -- admission of bound
orbits around a static object \cite{dadhich:2013}. Since for Einstein gravity potential goes as $1/r^{d-3}$,
bound orbits could exist only if $d-3<2$, the condition required for centrifugal potential to be able to
counter-balance gravitational pull. Thus for Einstein gravity, bound orbits can exist only in $d=4$ and
none else.

The situation however changes for pure Lovelock because for it potential goes as $1/r^\alpha$ where
$\alpha = (d-2N-1)/N$ \cite{pons:2012}. Clearly $\alpha<2$ always and hence bound orbits always exist in
pure Lovelock gravity in dimensions $d\geq 2N+2$. Then the question arises, is pure Lovelock static black hole
with potential $1/r^\alpha$, $\alpha = (d-2N-1)/N$ stable under scalar, vector and tensor perturbations? This
was the question addressed for pure Gauus-Bonnet black hole in six dimension \cite{Gannouji:2013eka}. It turns out that it
is in general unstable, however on inclusion of positive $\Lambda$, there does appear a parameter window for
mass and $\Lambda$ for which stability is achieved. Note that in this case potential falls off as $1/\sqrt{r}$
which is slower than $1/r$.

Further it turns out that pure Lovelock potential falls off exactly as four dimensional GR, $1/r$ in $d=3N+1$ \cite{chakraborty:2016}. In
all $d>3N+1$, it would fall sharper than $1/r$. The natural question that arises is that would pure Lovelock black hole
in $d=3N+1$ be stable? This is the question we wish to address in this paper. We shall show in particular that pure
GB black hole is stable in seven dimension. We shall also compute its quasinormal modes. This result will also be generalized to all dimensions $d\geq 3N+1$. The inference that emerges is that stability is determined by fall off potential whether it is slower or sharper than $1/r$. It is expected to be unstable for the former and stable for the latter irrespective of the pure Lovelock degree. Though it has been proven only for pure GB that black hole is unstable in $d=6$ where potential falls off as $1/\sqrt{r}$ \cite{Gannouji:2013eka}. In all dimensions $d< 3N+1$, potential falls off slower than $1/r$, black hole would be unstable.

Note that the two defining and distinguishing properties of pure Lovelock gravity are universalization of the kinematic property in all critical odd $d=2N+1$ dimensions and existence of bound orbits around a static object \cite{dadhich:2015}. If these are taken as the desirable and required properties for gravity in higher dimensions, pure Lovelock is the only right and proper gravitational equation in higher dimensions.

We should also mention another very desirable aspect of these theories. In fact, Lovelock gravity admits non-unique degenerate vacua \cite{Wheeler:1985nh,Wheeler:1985qd} which is not conducive for a supergravity extension of the theory. In that direction, various solutions have been proposed. For example, if we fine-tune coupling constants of the Lovelock polynomial as done in the consideration of dimensionally continued black holes \cite{Banados:1993ur}, it is possible to define a unique vacuum. 
It leads to the Chern-Simons and the Born-Infeld gravity in odd and even dimensions respectively \cite{Troncoso:1999pk}. The natural and elegant solution to have the unique vacuum is the pure Lovelock gravity. Pure Lovelock gravity has several interesting and remarkable features, see e.g. \cite{Dadhich:2015ivt} for a nice account.

The paper is organized as follows. In section II, we review black hole solution for pure Lovelock in critical dimension. In section III, we obtain stability of the black hole solution, in particular, we prove the stability of black holes under all type of perturbations. In section IV, we discuss the isospectrality of the perturbation modes. In section V, we obtain the QNM for scalar, vector and tensor perturbations for pure Gauss-Bonnet in $d=7$, before generalizing them to any critical dimension in section VI. Finally, we summarize our results.

\section{Pure-Lovelock in $d=3N+1$ dimension}

The Lovelock polynomial corresponds to the most general divergence free symmetric tensor constructed out of a metric and its first and second derivatives which gives a second order differential equation. The Lagrangian is given by
\begin{align}
\mathcal{L}=\sum_{k=0}^m c_k \mathcal{L}_{k}
\end{align}
where
\begin{align}
\mathcal{L}_{k}\equiv \frac{1}{2^k}\delta_{\alpha_1 \beta_1 \cdots \alpha_k \beta_k}^{\mu_1 \nu_1 \cdots \mu_k \nu_k} R^{\alpha_1 \beta_1}_{\mu_1 \nu_1}~^{\cdots}_{\cdots}R^{\alpha_k \beta_k}_{\mu_k \nu_k}
\end{align}
where $R^{\alpha_1 \beta_1}_{\mu_1 \nu_1}$ is the Riemann tensor in d-dimensions and $\delta_{\alpha_1 \beta_1 \cdots \alpha_k \beta_k}^{\mu_1 \nu_1 \cdots \mu_k \nu_k}$ is the generalized totally antisymmetric Kronecker delta. Pure Lovelock is the special case where the polynomial reduces to a monomial of order $N$, $\mathcal{L}=\mathcal{L}_{N}$.

Considering pure Lovelock, variation of the action gives
\begin{align}
\mathcal{G}_A^B=\delta_{A\alpha_1 \beta_1 \cdots \alpha_N \beta_N}^{B \mu_1 \nu_1 \cdots \mu_N \nu_N} R^{\alpha_1 \beta_1}_{\mu_1 \nu_1}~^{\cdots}_{\cdots}R^{\alpha_N \beta_N}_{\mu_N \nu_N}=0
\end{align}
There exist static exact solutions to this equation. For a static spherically symmetric spacetime of the following form,
\begin{align}
{\rm d}s^2 =-f(r)dt^2+\frac{dr^2}{f(r)}+r^2d\Omega_{d-2}^2
\end{align}
we have
\begin{align}
f(r)=1-\Bigl(\frac{r_s}{r}\Bigr)^{(d-2N-1)/N}
\end{align}
where $r_s$ is the location of the event horizon. Notice that $d=3N+1$ gives the Schwarzschild solution. In the rest of the paper, we will consider for each pure Lovelock of order $N$, the critical dimension corresponding, $d=3N+1$, which corresponds to Schwarzschild spacetime at any order $N$.

In the next section, we will study the stability of this spacetime and quasinormal modes (QNM) associated to the perturbations.

\section{Stability}
Stability of black holes has its root in the seminal work by Regge and Wheeler in 1957 \cite{Regge:1957td}. Perturbations around a given static spherically symmetric background can be decomposed into scalar and vector perturbations while a third type of perturbations emerges in dimensions larger than 4, known as tensor perturbations. While 4-dimensional stability has been thoroughly investigated by several authors in \cite{Vishveshwara:1970cc,Price:1971fb,Wald}, the extra dimension analysis is based on the formalism developed in \cite{Ishibashi:2003ap,Kodama:2003jz}. This formalism has been successfully used for many models of gravity beyond general relativity \cite{Takahashi:2010ye,Ganguly:2017ort}. The study of stability of black hole has important impact on the possible end state of a gravitational collapse. It is interesting to note that instability of black hole could also have a rich phenomenology, e.g. in the context of brane world models and black strings. The spacetime is plagued with the so-called Gregory-Laflamme instability leading to a naked singularity and thereby violating the cosmic censorship conjecture. But as suggested by the membrane paradigm, some duality could exist between Einstein and Navier-Stokes equations. In this context, it was suggested that Gregory-Laflamme instability could be mapped to effective fluid properties and the Plateau-Rayleigh instability \cite{Lehner:2011wc} or how a falling stream of fluid breaks up into smaller packets as the black string does. Also, in the context of asymptotically AdS spacetime with electric charge, instability can be related to superconducting phase transition of the dual theory on the boundary \cite{Hartnoll:2008vx}, see e.g. \cite{Aranguiz:2015voa} in the context of Lovelock.

In this direction, we consider a perturbation ($h_{\mu \nu}$) to the previous background solution described by the metric ($\bar g_{\mu \nu}$), $g_{\mu \nu}=\bar g_{\mu \nu}+h_{\mu \nu}$. The perturbation $h_{\mu \nu}$ can be decomposed into scalar, vector and tensor modes according to the symmetries of the background. For scalar perturbations, the metric in the Zerilli gauge takes the following form
\begin{align}
h_{\mu \nu}=\begin{pmatrix}
f H_0 & H_1 & \textbf{0} \\
H_1 & H_2/f & \textbf{0} \\
\textbf{0} & \textbf{0} & r^2 K \gamma_{ij}
\end{pmatrix}
\end{align}
where $\gamma_{ij}$ is the background metric for coordinates which are not ($t,r$). Each perturbation is decomposed in terms of scalar harmonics which satisfy $\bar\nabla_k \bar\nabla^k Y=-\gamma_s Y$ with $\gamma_s=l(l+d-3)$ and $\bar\nabla_k$ is the covariant derivative associated to the background metric. As shown in \cite{Takahashi:2010gz,Takahashi:2010ye}, all perturbations $(H_0,H_1,H_2,K)$ can be expressed by a single master function $\psi$ solution of the equation
\begin{align}
\frac{d^2\psi}{d r_*^{2}}+(\omega^2-V_s(r))\psi=0
\label{eq:Psis}
\end{align}
where $r_*$ is the tortoise coordinate, with $\omega$ comes the time dependence of the perturbation $e^{-i\omega t}$ and $V_s(r)$ is an effective potential for scalar perturbation
\begin{align}
V_s(r)=f(r)\frac{2\lambda^2(\lambda+1)r^3+6 \lambda^2M r^2+18\lambda M^2 r+18 M^3}{r^3(\lambda r+3 M)^2}
\end{align}
which we notice has exactly the same form as the Zerilli potential, with
\begin{align}
\lambda=\frac{(l+d-2)(l-1)}{d-2}
\end{align}
From which we can easily conclude that scalar perturbations are stable because $V_s(r)>0$ for $r>r_s$.

Similarly, for vector perturbations in the Regge-Wheeler gauge, we have
\begin{align}
h_{\mu \nu}=\begin{pmatrix}
0 & 0 & \textbf{v}_i \\
0 & 0 & \textbf{w}_i \\
\textbf{v}_i & \textbf{w}_i & \textbf{0}
\end{pmatrix}
\end{align}
with $\bar \nabla_k v^k=\bar \nabla_k w^k=0$ and each perturbation can be decomposed in terms of vector harmonics which satisfy the equation $\bar\nabla_k \bar\nabla^k Y_i=-\gamma_v Y_i$ with $\gamma_v=l(l+d-3)-1$. As for scalar perturbations, we can reduce the problem to a single function $\psi$ solution of
\begin{align}
\frac{d^2\psi}{d r_*^{2}}+(\omega^2-V_v(r))\psi=0
\label{eq:Psiv}
\end{align}
with
\begin{align}
V_v=f(r)\Bigl[\frac{(l+d-4)(l+1)}{(d-3)r^2}-\frac{6 M}{r^3} \Bigr]
\end{align}
which is also the simple generalization of the Regge-Wheeler equation to any dimension. Therefore we can also conclude that scalar perturbations decay with time because $V_v(r)>0$ for $r>r_s$.

Finally, for tensor perturbations, we have
\begin{align}
h_{\mu \nu}=\begin{pmatrix}
\textbf{0} & \textbf{0} \\
\textbf{0} & r^2\phi h_{ij}
\end{pmatrix}
\end{align}
with $h_{ij}$ the traceless, transverse tensor harmonics solution of $\bar\nabla_k \bar\nabla^k h_{ij}=-\gamma_t h_{ij}$ with $\gamma_t=l(l+d-3)-2$. Similarly, we can define a new function $\psi$ related to the perturbation $\phi$ solution of the equation
\begin{align}
\frac{d^2\psi}{d r_*^{2}}+(\omega^2-V_t(r))\psi=0
\end{align}
with
\begin{align}
V_t=f(r)\frac{2M}{r^3}
\label{eq:Vt}
\end{align}
which is a new perturbation inexistent in 4 dimensions. This perturbation is independent of the angular momentum $l$, they are no angular dependence. The perturbation oscillates similarly on a sphere of radius $r$ around the horizon. These perturbations are also stable because $V_t(r)>0$ for $r>r_s$.

Therefore for any critical dimension, $d=3N+1$, the Schwarzschild solution is always stable.

\section{Isospectrality}

An other way to study the stability of the black hole is to find the corresponding frequencies $\omega$. This problem can be seen as an eigenvalue problem
\begin{align}
\Bigl[-\frac{d^2}{d r_*^{2}}+V(r)\Bigr]\psi=\omega^2\psi
\end{align}
where the wavefunction $\psi$ is the eigenvector and $\omega$ the eigenvalue of the corresponding operator. These frequencies or eigenvalues of the previous operator are complex, and because we considered a time dependence of the following form $e^{-i\omega t}$, we should find that the imaginary part of these frequencies is negative to obtain a mode decaying with time
\begin{align}
e^{-i\omega t}=e^{-i\omega_R t}e^{\omega_I t}
\end{align}
where $\omega_R$ and $\omega_I$ are real and complex part of these modes known as quasinormal modes.

In 4 dimensions, it is well known that modes of Schwarzschild black hole are isospectral \cite{Chandrasekhar:1985kt}, the quasinormal modes are identical. This is an important result, specific to Schwarzschild black hole and does not hold for neutron stars or most of other gravitational theories. It can be shown that in general relativity, scalar potential $(V_s)$ and vector potential $(V_v)$ can be related to a same potential $W$
\begin{align}
W&=\frac{2M}{r^2}-\frac{3+2c}{3r}+\frac{c(3+2c)}{3 (3M+cr)}-\frac{c(c+1)}{3M}\label{eq:W}\\
\beta &= -\frac{c^2(c+1)^2}{9M^2}\\
c &=\frac{(l+2)(l-1)}{2}
\end{align}
The potentials can be obtained as
\begin{align}
V_s(r) &=W^2-f(r)\frac{{\rm d}W}{{\rm d}r}+\beta\\
V_v(r) &=W^2+f(r)\frac{{\rm d}W}{{\rm d}r}+\beta
\end{align}
If $\psi_v$ is an eigenfunction of the wavelike equation (\ref{eq:Psiv}), then the eigenfunction for the potential $V_s$ is given by
\begin{align}
\psi_s\propto \Bigl(W-f(r)\frac{{\rm d}}{\rm d r}\Bigr)\psi_v
\end{align}
and corresponds to the same eigenvalue $\omega$. Therefore, the quasinormal spectrum is the same for both perturbations.

In our case, we can show an almost isospectral behavior. All potentials can be derived from the same form of $W$ (\ref{eq:W}) and the potentials can be obtained in the following way
\begin{align}
V_s(r) &=W^2-f(r)\frac{{\rm d}W}{{\rm d}r}+\beta\\
V_v(r) &=W^2+f(r)\frac{{\rm d}W}{{\rm d}r}+\beta\\
V_t(r) &=W^2-f(r)\frac{{\rm d}W}{{\rm d}r}+\beta
\end{align}
but the spectrum of each operator is different because the constant $c$ differs for each type of perturbations
\begin{align}
&c=\frac{(l+d-2)(l-1)}{d-2}\,, ~~ &\text{for scalar perturbations} \nonumber\\
&c=\frac{(l+d-4)(l+1)}{2(d-3)}-1\,, ~~ &\text{for vector perturbations} \nonumber\\
&c=0\,, ~~ &\text{for tensor perturbations} \nonumber
\end{align}
In fact, the potentials are not obtained from the same potential $W$, because $c$ is different. In $d=4$, tensor modes do not exist and $c$ takes the same value for scalar and vector perturbations. We, therefore, recover the standard result. In any other critical dimension, the spectrum will be different.

\section{Quasinormal modes}

\begin{table*}
\small
\centering
\begin{tabular}{|c|c|c|c|c|c|cl}\hline
\multicolumn{6}{|c|}{}			\\
		\multicolumn{6}{|c|}{Scalar perturbations}			\\ \hline
		\multicolumn{6}{|c|}{$\ell = 2$}			\\ \hline
$n$	&		WKB method	&	Continued fraction  & Direct integration  & Schwarzschild  & Schwarzschild-Tangherlini	\\
	&			&	method &   & in $D=4$ &  in $D=7$	\\\hline
0	&	0.621586 - 0.174806 $i$	& 0.621745 - 0.174513 $i$	&	0.621745 - 0.174513 $i$ & 0.747415 - 0.177847 $i$ & 1.44794 - 0.46559 $i$\\
1	&	0.554145 - 0.546684 $i$	& 0.555450 - 0.542138 $i$	&	0.558945 - 0.546182 $i$ & 0.693431 - 0.547752 $i$ & 1.00452 - 1.49939 $i$\\
2	&	0.441637 - 0.974783 $i$	& 0.440166 - 0.97515 $i$ &	0.440006 - 0.969969 $i$ & 0.600099 - 0.957657 $i$ & 0.12018 - 2.94122 $i$\\
\hline
		\multicolumn{6}{|c|}{$\ell = 3$}			\\ \hline
$n$	&		WKB method	&	Continued fraction  & Direct integration  & Schwarzschild  & Schwarzschild-Tangherlini	\\
	&			&	method &   & in $D=4$ &  in $D=7$	\\\hline
0	&	0.952926 - 0.182173 $i$	& 0.952919 - 0.182185 $i$ &	0.952919 - 0.182186 $i$  & 1.198887 - 0.185406 $i$ & 2.23178 - 0.54084 $i$\\
1	&	0.910869 - 0.555971 $i$	& 0.909113 - 0.561265 $i$ &	0.907645 - 0.557741 $i$  & 1.165284 - 0.562585 $i$ & 1.93141 - 1.68185 $i$\\
2	&	0.835593 - 0.956598 $i$	& 0.836303 - 0.956543 $i$ &	0.800021 - 0.949965 $i$  & 1.10319 - 0.95811 $i$ & 1.33485 - 2.99002 $i$\\
\hline
\multicolumn{6}{|c|}{}			\\
		\multicolumn{6}{|c|}{Vector perturbations}			\\ \hline
		\multicolumn{6}{|c|}{$\ell = 2$}			\\ \hline
$n$	&		WKB method	&	Continued fraction  & Direct integration  & Schwarzschild  & Schwarzschild-Tangherlini	\\
	&			&	method &   & in $D=4$ &  in $D=7$	\\\hline
0	&		0.484860 - 0.178144 $i$	&	0.489874 - 0.171145  $i$  &  0.489877 - 0.171143  $i$	& 0.747343 - 0.177925 $i$  &  1.96718 - 0.60766 $i$\\
1	&		0.392473 - 0.578251 $i$	&	0.395578 - 0.540791  $i$   &  0.375665 - 0.526008  $i$	&	0.693422 - 0.547830 $i$  &  1.52058 - 1.90469 $i$\\
2	&		0.258772 - 1.083567 $i$	&	0.211030 - 1.006590  $i$   &  0.249988 - 1.000020  $i$ & 0.602107 -  0.956554 $i$  &  0.57082 - 3.61342 $i$\\\hline
		\multicolumn{6}{|c|}{$\ell = 3$}			\\ \hline
$n$	&		WKB method	&	Continued fraction  & Direct integration  & Schwarzschild  & Schwarzschild-Tangherlini	\\
	&			&	method &   & in $D=4$ &  in $D=7$	\\\hline
0	&	0.747239 - 0.177782 $i$  & 0.747343 - 0.177925 $i$  & 0.747343 - 0.177925 $i$   & 1.198887 - 0.185406 $i$  &  2.64217 - 0.60118 $i$\\
1	&	0.692593 - 0.546960 $i$	& 0.693422 - 0.547830 $i$  & 0.695469 - 0.545779 $i$   & 1.165284 - 0.562581 $i$  &  2.30384 - 1.84962 $i$\\
2	&	0.597040 - 0.955120 $i$	& 0.602107 - 0.956554 $i$  & 0.599970 - 0.950054 $i$   & 1.10319 - 0.95809 $i$      &  1.58364 - 3.29736 $i$\\
\hline
\multicolumn{6}{|c|}{}			\\
		\multicolumn{6}{|c|}{Tensor perturbations}			\\ \hline
		\multicolumn{6}{|c|}{For any $\ell$}			\\ \hline
$n$	&		WKB method	&	Continued fraction  & Direct integration  & Schwarzschild  & Schwarzschild-Tangherlini	\\
	&			&	method &   & in $D=4$ &  in $D=7$	($\ell = 2$)\\\hline
0	&	 0.220934 - 0.201633 $i$	& 0.220910 - 0.209791 $i$ & 0.225108 - 0.214605 $i$ & -   &  2.49710 - 0.63234 $i$ \\
1	&	 0.178058 - 0.689058 $i$	& 0.172234 - 0.696105 $i$ & 0.148482 - 0.678334 $i$ & -   &  2.13340 - 1.98070 $i$ \\
2	&	 0.383649 - 0.952812 $i$	& 0.381050 - 0.952612 $i$ & 0.380012 - 0.95002 $i$ & -   &  1.37830 - 3.64177 $i$ \\\hline \hline
\end{tabular}
\caption{QNM for scalar, vector and tensor perturbations for $\ell=(2, 3)$ and $n=(0, 1, 2)$ by using 3 different methods. Also we added for comparison, the QNM for Schwarzschild in $D=4$ and Schwarzschild-Tangherlini in $D=7$.}
\label{tableQNM}
\end{table*}

As we have shown previously, the problem can be reduced to finding eigenvalues of some operator. These frequencies, or quasinormal modes, gives us information on the proper oscillations of perturbed spacetime, the characteristic ``ringing'' of black holes. In this section, we will calculate the spectrum $(\omega)$ for each type of perturbations, solution of the equation
\begin{align}
\frac{d^2\psi}{d r_*^{2}}+\Bigl(\omega^2-V(r)\Bigr)\psi=0
\label{eq:master}
\end{align}
where $V(r)$ is the potential for scalar, vector or tensor perturbations. We also need to impose boundary conditions, at the horizon, the perturbation should be ingoing $(e^{-i\omega r_*})$ and the perturbation should be outgoing at infinity $(e^{+i\omega r_*})$. Imposing these boundary conditions, only an infinite (labeled by a parameter $n$) but discrete values of $\omega$ solve the eq. (\ref{eq:master}), known as the quasinormal modes (QNM). They are the modes at which the spacetime oscillates and because we have dispersion of the gravitational waves (ingoing at horizon and outgoing at infinity), these modes are complex and therefore decay in time. The imaginary part should be negative because the black hole is stable as seen in the previous section. All the modes $\omega$ will be given in units $r_s/c=1$

In this section, we will focus on the dimension 7, which is our first new critical dimension after standard Einstein gravity in $d=4$. In this case, the action is given by pure Gauss-Bonnet
\begin{align}
\mathcal{S}=\int {\rm d}^7x \sqrt{-g} \Bigl[R^2-4R_{\mu \nu}R^{\mu \nu}+R_{\mu \nu \rho \sigma} R^{\mu \nu \rho \sigma}\Bigr]
\end{align}
For each type of perturbations, we will fix the angular momentum $(\ell=2, 3)$ and calculate the first harmonic or fundamental mode $(n=0)$, the first overtone or second harmonic ($n=1$) and finally the second overtone or third harmonic $(n=2)$. For that, we will perform the calculations using 3 different methods, the WKB method\cite{Schutz:1985zz,Iyer:1986np} at sixth order\cite{Konoplya:2003dd,Konoplya:2003ii}, the continued fraction method \cite{Leaver:1985ax} and by direct integration. The quasinormal modes are compared to Schwarzschild spacetime in $d=4$ and to the Schwarzschild-Tangherlini solution \cite{Tangherlini:1963bw} in $d=7$. Therefore, we perform the comparison to same background while the action is modified and to same dimension while the action is changed.

The Schwarzschild-Tangherlini solution in 7 dimensions is derived from Einstein action and gives $f(r)=1-(r_s/r)^4$. Using exactly the same methodology, the new equation for perturbations will have exactly the same structure than eq. (\ref{eq:master}) with following potentials for scalar, vector and tensor perturbations respectively
\begin{align}
V_s(r)&=\Bigl[1-\Bigl(\frac{r_s}{r}\Bigr)^{4}\Bigr]\Bigl[\frac{5 r^{12} \lambda^2 (4 \lambda+7) + 9 r^4 r_s^8 (26 \lambda+3)}{4 r^6 (\lambda r^4 +3 r_s^4 )^2}\nonumber\\
&\qquad- \frac{15 r^8 r_s^4 \lambda ( \lambda+18) +225 r_s^{12} }{4 r^6 (\lambda r^4 +3 r_s^4 )^2}\Bigr]\\
V_v(r)&=\Bigl[1-\Bigl(\frac{r_s}{r}\Bigr)^{4}\Bigr]\Bigl[\frac{(2l+5)(2l+3)}{4 r^2}-\frac{75 r_s^4}{4 r^6}\Bigr]\\
V_t(r)&=\Bigl[1-\Bigl(\frac{r_s}{r}\Bigr)^{4}\Bigr]\Bigl[\frac{(2l+5)(2l+3)}{4 r^2}+\frac{25 r_s^4}{4 r^6}\Bigr]
\end{align}

Notice from Table \ref{tableQNM} that for all modes $\omega_I<0$ and therefore the solution is stable. It is also interesting that ringdown signal vanishes exponentially with a characteristic time $\tau=1/\text{Im}(\omega)$ which is very similar between the Schwarzschild solution in 7 dimensions and the Schwarzschild spacetime in 4 dimensions. The dimension affects very little the characteristic time, while it is very different for similar dimension $D=7$ and within different solution, Schwarzschild-Tangherlini. The characteristic time of decay of the perturbations is faster for Schwarzschild-Tangherlini background where the gravitational potential falls faster to zero ($1/r^4$). The characteristic time at which the perturbation decay depends mostly on the background potential.

On the contrary, the frequency at which these modes oscillate ($f=\text{Re}(\omega)/2\pi$) is very different for the 3 spacetimes considered but again the analogue Schwarzschild spacetime has a frequency of mode oscillation closer to Schwarzschild spacetime in $D=4$ than Schwarzschild-Tangherlini in $D=7$. We conclude that this parameter is more sensitive to the theory considered and to the dimension.

\section{Higher dimensions}

\begin{figure}[h]
\begin{centering}
\includegraphics[scale=.5]{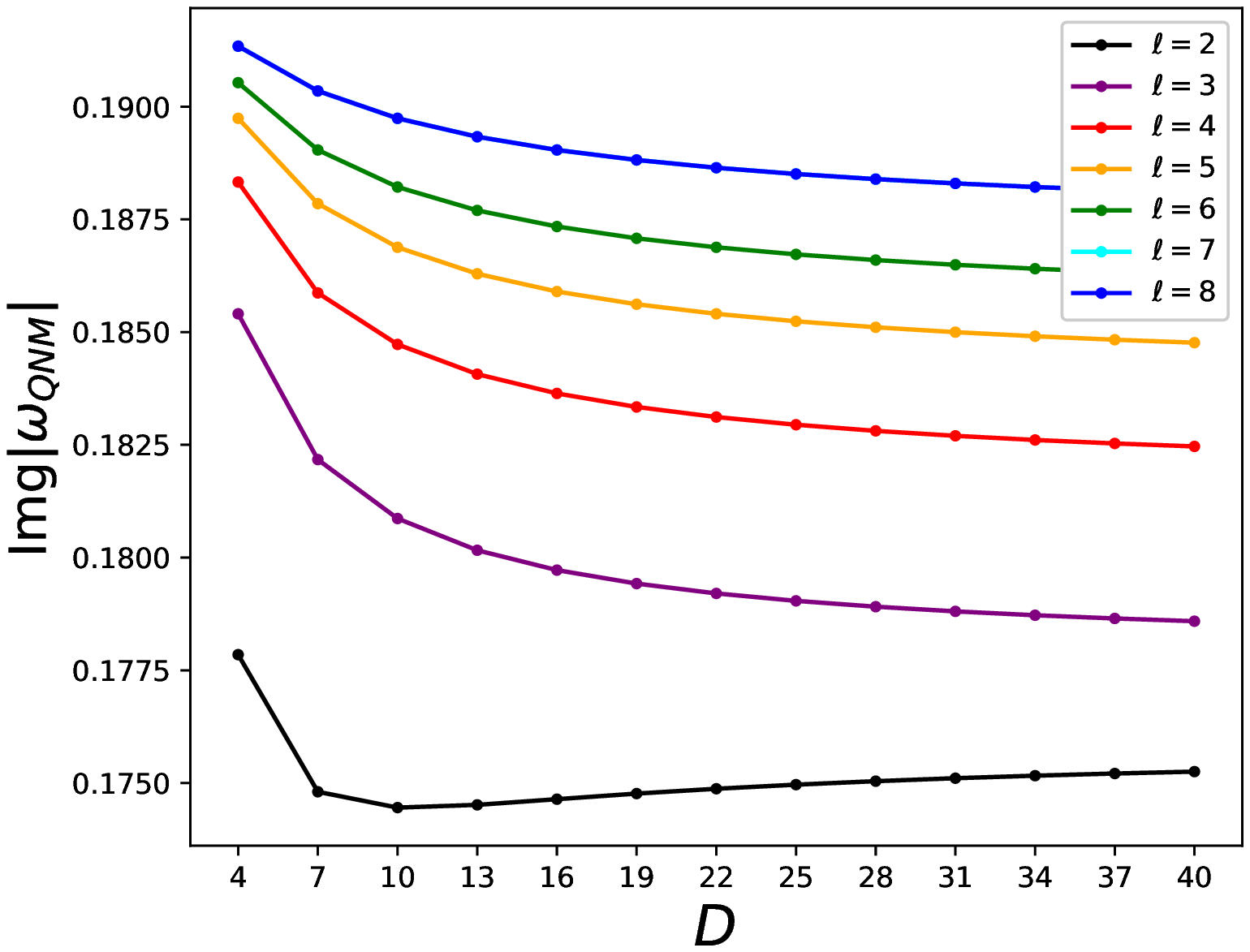}
\includegraphics[scale=.5]{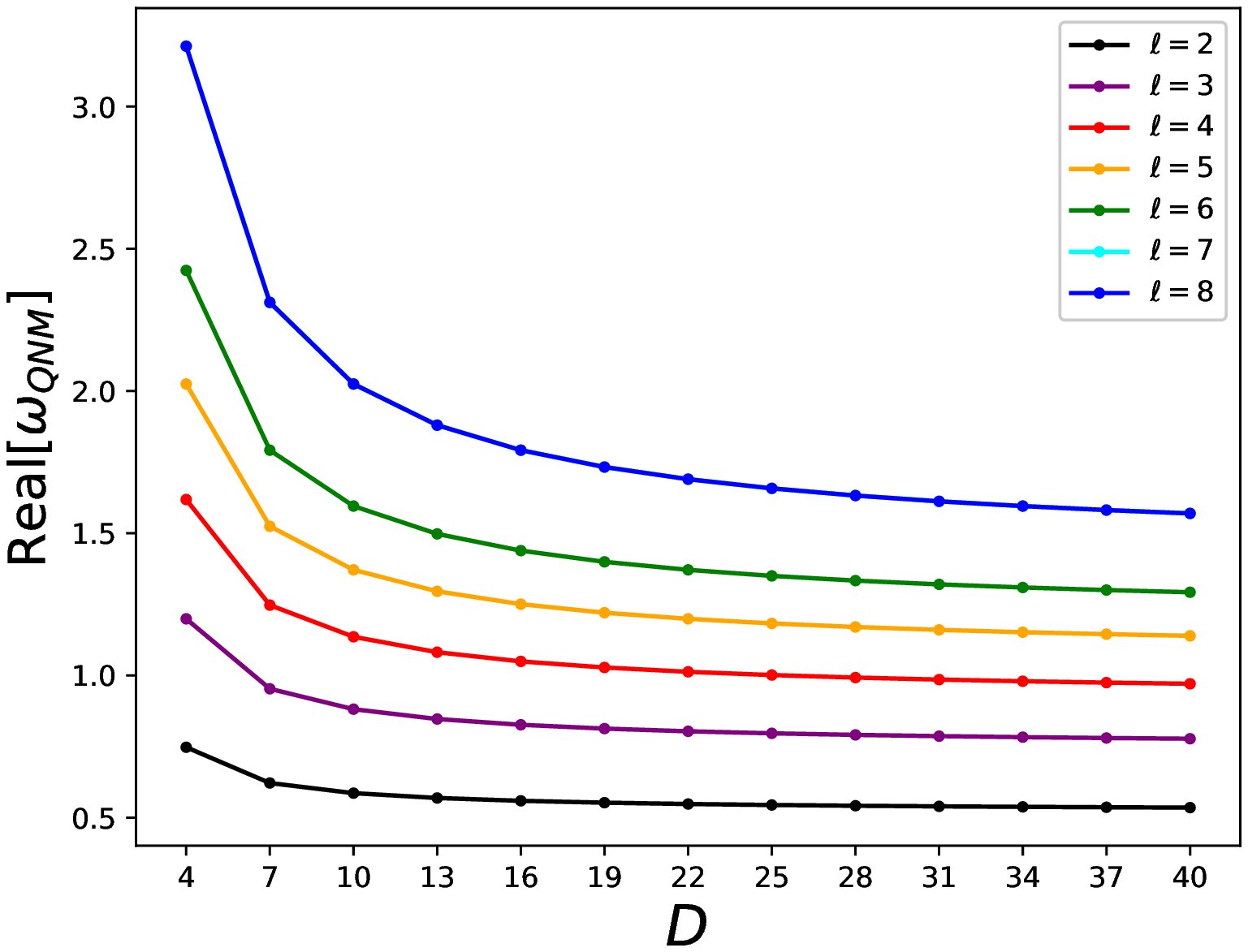}
\par\end{centering}
\caption{Imaginary part of QNM (upper) and real part of QNM (lower) for scalar perturbations in all critical dimensions from $D=4$ to $D=40$ using WKB method of order 6 for the fundamental tone $n=0$ and angular momentum $l=2, ..., 8$.}
\label{figScalar}
\end{figure}
\begin{figure}[h]
\begin{centering}
\includegraphics[scale=.5]{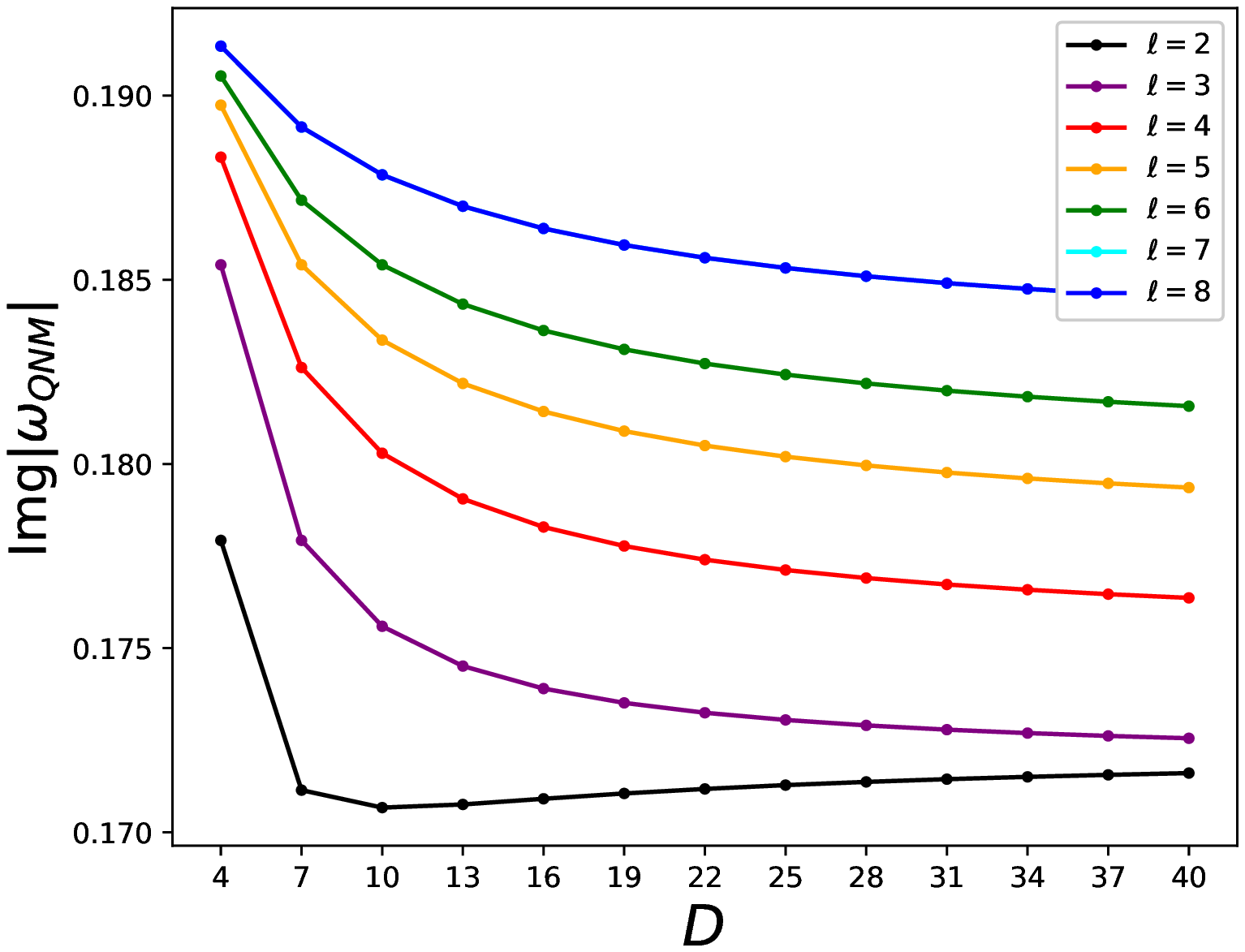}
\includegraphics[scale=.5]{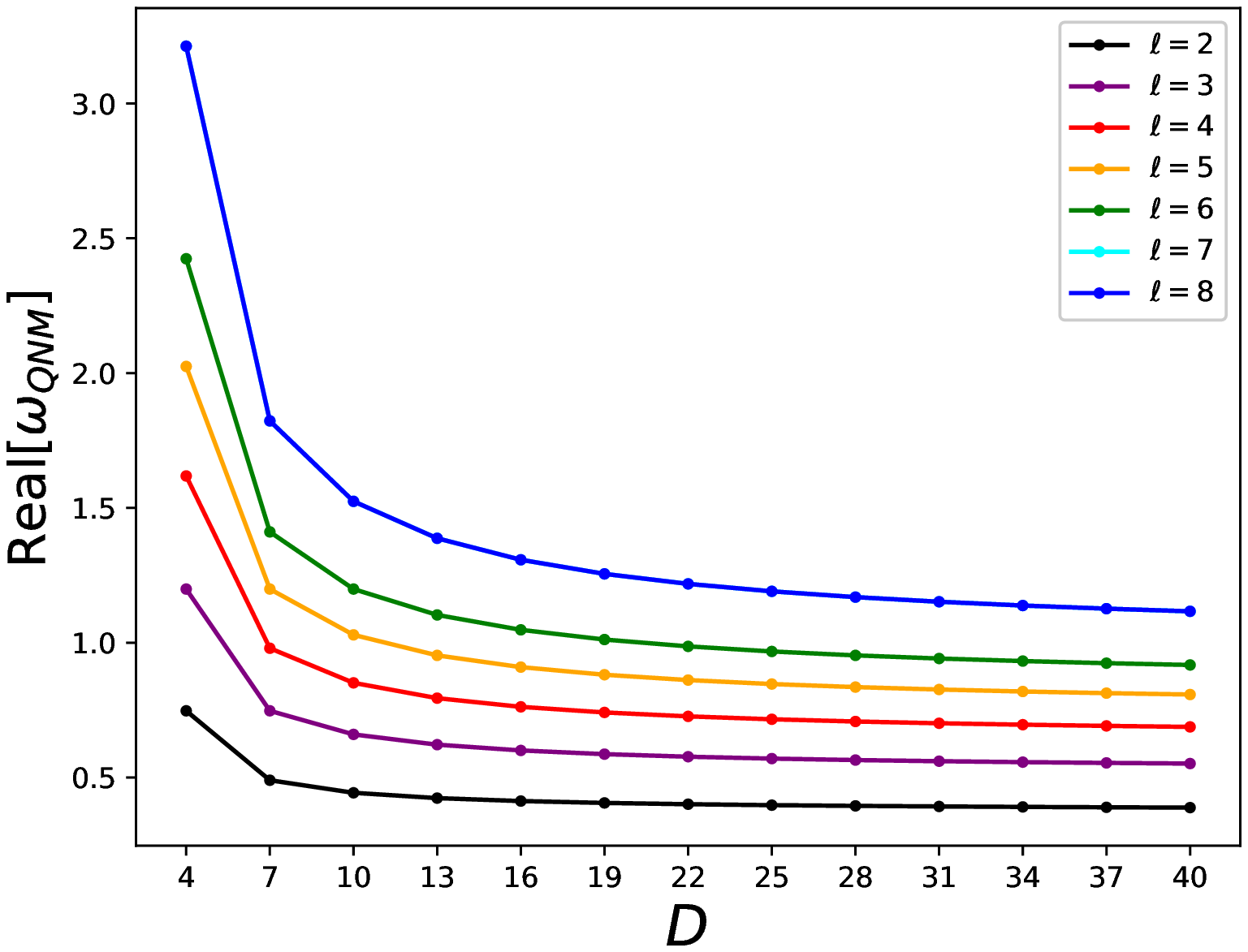}
\par\end{centering}
\caption{Imaginary part of QNM (upper) and real part of QNM (lower) for vector perturbations in all critical dimensions from $D=4$ to $D=40$ using Leaver's method for the fundamental tone $n=0$ and angular momentum $l=2, ..., 8$.}
\label{figVector}
\end{figure}

In this section, we generalize the previous results to any critical dimensions, Lovelock of order $N$ in dimension $d=3N+1$. As we can see from eq.(\ref{eq:Vt}), tensor modes do not dependent on the dimension, they oscillate similarly in any critical dimension. On the contrary, scalar and vector perturbations depend on the dimension.

We see from Figs.(\ref{figScalar},\ref{figVector}) similar results than we found in the previous section. The characteristic time at which the perturbations die is changing very little between all these models. Dimension does not affect the characteristic time of the ringdown, only the background gravitational potential affects it. Notice that for $l=2$, the characteristic time ($1/\text{Im}(\omega)$) becomes maximum for $d=10$ for scalar and vector perturbations.

On the contrary, the frequency of oscillation is changing much more. It depends on the type of the action, and therefore on the dimension of the spacetime. The larger the dimension, the lower the frequency. Notice also that larger angular momentum $l$ increases these effects. Finally, we can notice that all values settle to some asymptotic value. This is due to the fact that $V_s$ and $V_v$ have well defined limit when $d\rightarrow \infty$, e.g. we have
\begin{align}
V_v=f(r)\Bigl[\frac{l+1}{r^2}-\frac{6M}{r^2}\Bigr]+\mathcal{O}(\frac{1}{d})
\end{align}
and therefore the results of Figs.(\ref{figScalar},\ref{figVector}) could be obtained analytically by an $1/d$ expansion in higher d-dimensional spacetime. 

\section{Conclusions}

We have shown that for any dimension $d=3N+1$, Schwarzschild spacetime is stable in the corresponding $N$th order pure Lovelock gravity. And it is an exact solution of pure Lovelock equation in $d=3N+1$. We have studied stability of this static spherically symmetric spacetime for  linear perturbations. We have derived perturbation equation and shown that isospectrality holds only in $d=4$, in any other dimension, scalar, vector and tensor perturbations oscillate with different frequencies. We have also studied QNM for all critical dimensions from $d=4$ to $d=40$ and we found that tensor perturbations are isotropic and do not depend on the dimension of the spacetime, while scalar and vector decay to zero with a characteristic time which depends weakly on dimension. This is because perturbations depend only on the gravitational potential while frequency of oscillation of these perturbations does depend on dimension. The larger the dimension, the lower is the frequency. We have also shown that all QNM's converge to an asymptotic value for large dimension, because the potentials $(V_s,V_v)$ have well defined limit for large $d$, and therefore could be tackled analytically using $1/d$ expansion. 

Pure Lovelock gravity is dynamical in $d\geq 2N+2$ however static black hole is stable only in dimension $d\geq 3N+1$. This means in dimensions $2N+2 \leq d < 3N+1$, black hole would be unstable. However for Einstein gravity $N=1$ the two limits coincide. For $N>1$, there would always be a dimension window of instability for pure Lovelock black holes. It would be interesting to study in a future work if third order Lovelock, $N=3$, is unstable in dimension $d=8,9$.

\section*{Acknowledgments}
R.G. is supported by Fondecyt project No 1171384. ND warmly acknowledges Albert Einstein Institute, Golm for the continued short term summer visits.


\end{document}